 \documentclass[aps,pra,twocolumn,groupedaddress,amsmath,amssymb,showpacs]{revtex4}
\usepackage{graphicx}
\usepackage{amsmath}
\usepackage{dcolumn}
\usepackage{bm}

\newcommand{\tr}{\mathrm{Tr}}
\newcommand{\inp}{\mathrm{in}}
\newcommand{\out}{\mathrm{out}}

\begin{document}
\title{Maximally Entangled Mixed-State Generation via Local Operations}
\author{A. Aiello}
\author{G. Puentes}
\author{D. Voigt}
\author{J.P. Woerdman}
\affiliation{Huygens Laboratory, Leiden University\\
P.O.\ Box 9504, 2300 RA Leiden, The Netherlands}
\begin{abstract}
We present a  general theoretical method to generate maximally
entangled mixed states  of a pair of photons initially prepared in
the singlet polarization state. This method requires only local
operations upon a single photon of the pair and exploits spatial
degrees of freedom to induce decoherence. We report also
experimental confirmation of these theoretical results.
\end{abstract}

 \maketitle
%
%
%
%
%
%
\section{Introduction}
Entanglement is perhaps the most puzzling feature of quantum
mechanics and in the last two decades it became the key resource
in quantum information processing \cite{NielsenBook}.
Entangled qubits prepared in pure, maximally entangled states are
required by many quantum-information processes. However, in a
mundane world, a pure maximally entangled state is an idealization
as, e.g., a plane wave in classical optics. In fact, interaction
of qubits with the environment leads to decoherence that may cause
a pure entangled state to become less pure (mixed) and less
entangled.
Thus, any \emph{realistic} quantum-communication/computation
protocol must cope with entangled mixed states and it is desirable
to attain the maximum amount of entanglement for a given degree of
mixedness. States that fulfill this condition
 are called maximally entangled mixed states
 (MEMS) and, recently, they have been the subject of several papers
 (see, e.g., \cite{Peters,Barbieri} and references therein).
In this Article we propose a new method to create MEMS from a pair
of photons initially prepared in the singlet polarization state.

Kwiat and coworkers \cite{Peters} were the first to achieve MEMS
 using photon pairs from spontaneous parametric down
conversion (SPDC). They induced decoherence in SPDC pairs
initially prepared in a pure entangled state  by coupling
polarization and time degrees of freedom of the photons. At the
same time, a somewhat different scheme was used by De Martini and
coworkers \cite{Barbieri} who instead used the spatial degrees of
freedom of SPDC photons to induce decoherence. However, both the
Kwiat and the De Martini method require operations on \emph{both}
photons of the SPDC pair. On the contrary, our technique has the
advantage to require only \emph{local} operations upon one of the
two photons.

This Article is structured as follows: In the first part of Sec.
II we show the relation existing between a one-qubit quantum map
and a classical-optics setup on the laboratory bench. In the
second part of Sec. II, we exploit this knowledge to design a
simple linear-optical set-up to generate MEMS from a pair of
photons via local operations and postselection.  Then,  in Sec.
III we provide an experimental demonstration of our method, using
entangled photons from parametric down-conversion. Finally, we
draw our conclusions in Sec. IV.
\section{Theory}
We begin by giving  a brief description of the connection between
classical polarization optics and quantum mechanics of qubits, as
recently put forward by several authors
\cite{Brunner03,Aiello061}.
Most textbooks on classical optics introduce the concept of
polarized and unpolarized light  with the help of the Jones and
Stokes-Mueller calculi, respectively \cite{Damask}. In these
calculi, the description of classical polarization of light is
formally identical to the quantum description of pure and mixed
states of two-level systems, respectively \cite{Iso}.
Mathematically speaking, there is an isomorphism between the
quantum density matrix $\rho$ describing a qubit and the classical
\emph{coherency matrix} $J$ \cite{BornWolf} describing
polarization of a beam of light: $\rho \sim J/\tr J$.
$J$ is an Hermitean, positive semidefinite $2 \times 2$ matrix, as
 is $\rho$.
%
%
A classical linear optical process (as, e.g., the passage of a
beam of light through an optical device), can be described by a $4
\times 4$ complex-valued matrix $\mathcal{M}$ such that
$(J_\mathrm{out})_{ij} = \mathcal{M}_{ij,kl}(J_\mathrm{in})_{kl}$,
where, from now on, we adopt the
 convention that summation over repeated Latin indices is
understood. Moreover, we assume that all Latin indices
$i,j,k,l,m,n, \ldots$ take the values $0$ and $1$, while Greek
indices $\alpha, \beta, \dots $ take the values $0,1,2,3$.
In polarization optics one usually deals with the real-valued
Mueller matrix $M$ which is connected to $\mathcal{M}$ via a
unitary transformation $\Lambda: M = \Lambda^\dagger \mathcal{M}
\Lambda$ \cite{AielloMath}.
The  matrix $M$ is often written as \cite{Lu96}
\begin{equation}\label{eq30}
M = \left(%
\begin{array}{cc}
  m_{00} & \mathbf{d}^T \\
  \mathbf{p} & W \\
\end{array}%
\right),
\end{equation}
where $(\mathbf{p}, \mathbf{d})\in \mathbb{R}^3$, are known as the
\emph{polarizance vector} and the \emph{diattenuation vector}
(superscript $T$ indicates transposition), respectively. Note that
$\mathbf{d}$ is nonzero only for dichroic media, namely media that
induce polarization-dependent losses (PDL) \cite{Damask}.
$W$ is a $3 \times 3$ real-valued matrix.
It should be noticed that if we choose $m_{00}=1$ (this can be
always done since it amounts to a trivial
\emph{polarization-independent} renormalization), the Mueller
matrix of a non-dichroic optical element ($\mathbf{d} =
\mathbf{0}$), is formally identical to a non-unital,
trace-preserving, one-qubit quantum map (also called channel)
\cite{Ruskai}. If also $\mathbf{p}=\mathbf{0}$ (pure depolarizers
and pure retarders  \cite{Damask}), then $M$ becomes identical to
a unital, one-qubit channel \cite{NielsenBook}.
It is not difficult to show that any linear optical device that
can be represented by $\mathcal{M}$ (or $M$), can also be
described by a set of at most four distinct optical elements in
parallel as $\mathcal{M} = \sum_{\alpha } \lambda_\alpha T_\alpha
\otimes T_\alpha^*$, where the four $2 \times 2$ \emph{Jones}
  matrices $T_\alpha$, represent four different
non-depolarizing optical elements and $\lambda_\alpha \geq 0$
\cite{Anderson94,AielloMath} .
From the results above it readily follows that the most general
operation that a linear optical element can perform upon a beam of
light can be written as $J_\inp \rightarrow J_\out = \sum_{\alpha}
\lambda_\alpha T_\alpha J_\inp T_\alpha^\dagger$.
Since $\lambda_\alpha \geq 0$, the previous equation is formally
identical to the Kraus form \cite{NielsenBook} of a completely
positive one-qubit quantum map $\mathcal{E}$. Therefore, if a
single photon encoding a polarization qubit passes through an
optical device classically described by the Mueller matrix
$\mathcal{M} = \sum_{\alpha } \lambda_\alpha T_\alpha \otimes
T_\alpha^*$, its initial state $\rho_\inp$ will be transformed
according to $\rho_\inp \rightarrow \rho_\out \propto \sum_\alpha
\lambda_\alpha T_\alpha \rho_\inp T_\alpha^\dagger$.

Now that we have learned how to associate a quantum map to a set
of at most four optical elements, we can apply this knowledge to
design a simple optical scheme suitable for MEMS production.
Suppose to have
 two qubits (encoded in the polarization degrees of
freedom of two SPDC photons, say $A$ and $B$), initially prepared
in the state $\rho: \rho = \rho_{ij,kl} |ij\rangle \langle kl|
\doteq  \rho_{ik,jl}^R |i\rangle \langle k| \otimes |j\rangle
\langle l| $. Superscript $R$ indicates \emph{reshuffling}
\cite{Zico} of the indices: $\rho_{ik,jl}^R \equiv \rho_{ij,kl}$.
Following Ziman and Bu\v{z}ek \cite{Ziman1} we assume that $\rho$
is transformed under the action of the most general \emph{local}
(that is, acting upon a single qubit) linear map $\mathcal{E}
\otimes \mathcal{I}$ into the state
\begin{equation}\label{eq50}
\rho_\mathcal{E} = \mathcal{E} \otimes \mathcal{I}[\rho] \propto
\sum_{\alpha=0}^3 \lambda_\alpha   T_\alpha  \otimes I  \, \rho \,
T_\alpha^\dagger \otimes I .
\end{equation}
By writing explicitly Eq. (\ref{eq50}) in the two-qubit basis $\{|
ij\rangle \equiv |i \rangle \otimes |j \rangle \}$, it is
straightforward to obtain $(\rho_\mathcal{E})_{ij,kl} \propto
\sum_\alpha \lambda_\alpha \rho^R_{mn,jl} (T_\alpha)_{im}
(T^*_\alpha)_{kn}$. Then, from the definition of $\mathcal{M}$ it
easily follows that $ (\rho_\mathcal{E})_{ij,kl} \propto (
\mathcal{M} \rho^R) _{ik,jl}$. By reshuffling $\rho_\mathcal{E}$,
this last result can be written in matrix form as
$\rho_\mathcal{E}^R \propto  \mathcal{M} \rho^R$ which displays
the very simple relation existing between the \emph{classical}
Mueller matrix $\mathcal{M}$ and the \emph{quantum} state
$\rho_\mathcal{E}$. Via a direct calculation, it is possible to
show that if $\rho$ represents two qubits in the singlet state
$\rho_s = \frac{1}{4}(I \otimes I - \sigma_x \otimes \sigma_x -
\sigma_y \otimes \sigma_y - \sigma_z\otimes \sigma_z)$
\cite{Pauli}, then the proportionality symbol in the last equation
above can be substituted with the equality symbol:
$\rho_\mathcal{E}^R = \mathcal{M} \rho^R_s$.
 Note that this
pleasant property is true only for the singlet state. However, if
the initial state $\rho$ is different from the singlet one, then
$\mathcal{M}$  must be simply renormalized by imposing $\tr
(\mathcal{M} \rho^R)=1$ .

Now, suppose that we have an experimental setup producing pairs of
SPDC photons in the singlet state $\rho_s$, and we want to
transform $\rho_s$ into the target state $\rho_\mathcal{T}$ via a
local map $\mathcal{T} \otimes \mathcal{I}: \, \rho_s \rightarrow
\rho_\mathcal{T} = (\mathcal{M}_\mathcal{T} \rho_s^R)^R$. All we
have to do is first to invert the latter equation to obtain
\begin{equation}\label{eq70}
\mathcal{M}_\mathcal{T} =  \rho_\mathcal{T}^R (\rho_s^R)^{-1},
\end{equation}
and then to decompose $\mathcal{M}_\mathcal{T}$ as
$\mathcal{M}_\mathcal{T} = \sum_{\alpha } \lambda_\alpha T_\alpha
\otimes T_\alpha^*$. Thus, we get the (at most four) Jones
matrices $ T_\alpha$ representing the optical elements necessary
to implement the desired transformation.
This is the main theoretical result of this Article.
Our technique is very straightforward and we shall demonstrate its
 feasibility later, by applying it  to  design an
optical setup devoted to MEMS generation. However, at this moment,
some caveats are in order. To make $\mathcal{M}_\mathcal{T}$  a
physically realizable Mueller matrix, its associated matrix
$H_\mathcal{T}$ should be positive semidefinite \cite{NotaH}.
 If
this is not the case, then the transformation $\rho \rightarrow
\rho_\mathcal{T}$ cannot be implemented via local operations. For
example, it is easy to see that if the initial state is a Werner
state $\rho_W = p\rho_s + \frac{1-p}{4}I, \; (0\leq p \leq 1)$ and
the target state is the singlet $\rho_\mathcal{T}=\rho_s$, then
such operation (known as \emph{concentration} \cite{Thew01R})
cannot be physically implemented by a local setup since
$H_\mathcal{T}$ has three degenerate negative eigenvalues. Another
caveat comes from the no-signalling constraint. Since
$\mathcal{M}_\mathcal{T}$ describes a local device operating only
upon photon $A$, a second observer watching at photon $B$ cannot
distinguish the initial state $\rho_s$ from the transformed state
$\rho_\mathcal{T}$, that is: $\rho^B = \tr_A (\rho_s) = \tr_A
(\rho_\mathcal{T})$. This condition requires the one-qubit map
$\mathcal{T}$ to be trace-preserving: $\sum_\alpha \lambda_\alpha
T_\alpha^\dagger T_\alpha = I$. From Eq. (\ref{eq30}), a
straightforward calculation shows that such condition cannot be
fulfilled if $\mathbf{d} \neq \mathbf{0}$, that is if the device
implementing $\mathcal{T}$ contains dichroic (or PDL) elements.
PDL is important in many commonly used optical devices as
polarizers, circulators, isolators, etc., \cite{Damask}. Within
the framework of quantum information theory, all these
\emph{physical} devices may be represented by
``\emph{unphysical}'' one-qubit maps $\mathcal{T}$ that violate
the no-signalling condition. This apparent paradox disappears if
one allows causal classical communications between observers who
actually measure and reconstruct the target state
$\rho_\mathcal{T}$ generated by the ``unphysical'' local map
$\mathcal{T} \otimes \mathcal{I}$ \cite{Aiello062}.
In fact, in coincidence measurements (required to reconstruct
$\rho_\mathcal{T}$), classical (as opposed to quantum) signalling
between the two observers is necessary to allow them to compare
their own experimental results and select from the raw data the
coincidence counts.
In other words, a coincidence measurement post-selects only those
photons that have not been absorbed by the PDL element
\cite{Brunner03}.
\begin{figure}[!htr]
\includegraphics[angle=0,width=8truecm]{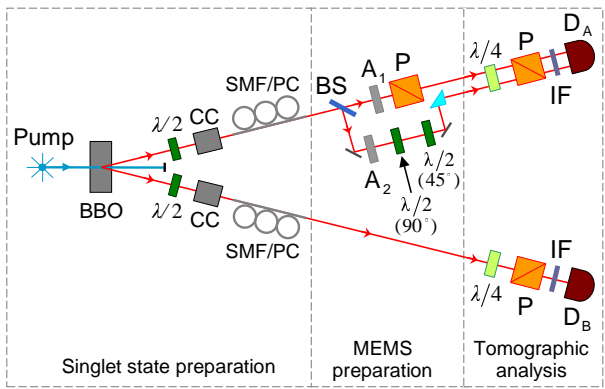}
\caption{\label{fig:A} (color online) Layout of the experimental
setup. The two-path optical device acts only on photon $A$.
Detectors $\mathrm{\textsf{D}}_\mathrm{\textsf{A}}$ and
$\mathrm{\textsf{D}}_\mathrm{\textsf{B}}$ perform coincidence
measurements.}
\end{figure}

With these caveats in mind,  we come to the experimental
validation of our method. We choose to generate MEMS I states
\cite{NoteMEMS}, represented by the density matrix
$\rho_\mathrm{I} = p | \phi_+ \rangle \langle \phi_+ | + (1-p)|01
\rangle \langle 01|$, where  $| \phi_+ \rangle  = (|00 \rangle +
|11 \rangle)/\sqrt{2}$ and $(2/3 \leq p \leq 1)$.
By varying the parameter $p$,  the entanglement and mixedness of
the state $\rho_\mathrm{I}$ change. Here, we use the linear
entropy $S_L$ \cite{Bose01R} and the tangle $T$, namely, the
concurrence squared \cite{Wootters98}, to quantify the degree of
mixedness and of entanglement, respectively. They are defined as
$S_L(\rho) = \frac{4}{3}[1 - \tr (\rho^2)]$, and  $T(\rho) =
[\max\{0 , \sqrt{\lambda_0} - \sqrt{\lambda_1} -\sqrt{\lambda_2}
-\sqrt{\lambda_3}\}]^2$, where $\lambda_0 \geq
\lambda_1\geq\lambda_2\geq \lambda_3 \geq 0$ are the eigenvalues
of $\rho (\sigma_y \otimes \sigma_y) \rho^* (\sigma_y \otimes
\sigma_y)$.
After applying Eq. (\ref{eq70}) with $\rho_\mathcal{T} =
\rho_\mathrm{I}$, a straightforward calculation shows that there
are only two non-zero terms in the decomposition of
$\mathcal{M}_\mathcal{T}$,
namely $\{\lambda_0 = 2(1-p) , \lambda_1 = p \}$, $ \{ T_0 =  \left(%
\begin{array}{cc}
  1 & 0 \\
  0 & 0 \\
\end{array}%
\right), T_1 = \left(%
\begin{array}{cc}
  0 & -1 \\
  1 & 0 \\
\end{array}%
\right)\}$. In physical terms, $T_0$ is a polarizer and $T_1$ is a
$90^\circ$ polarization rotator. The two eigenvalues $\{\lambda_0,
\lambda_1 \}$ give the relative intensity in the two arms of the
device and are physically realized by intensity attenuators.
\section{Experiment}
Our experimental set-up is shown in Fig.1. Its first part
(\textsf{Singlet state preparation}) comprises a Krypton-ion laser
at 413.1~nm that pumps a 1-mm thick $\beta-\mathrm{Ba}
\mathrm{B}_2 \mathrm{O}_4$ (\textsf{BBO}) crystal, where
polarization-entangled photon pairs at wavelength 826.2~nm are
created by SPDC in degenerate type II phase-matching configuration
\cite{Kwiat95}. Single-mode fibers (\textsf{SMF}) are used as
spatial filters to assure that the initial two-photon state is in
a single transverse mode. Spurious birefringence along the fibers
is compensated by suitably oriented polarization controllers
(\textsf{PC}) \cite{Puentes061}. In addition, total retardation
introduced by the fibers and walk-off effects at the \textsf{BBO}
crystal are compensated by compensating crystals (\textsf{CC}:
0.5-mm thick \textsf{BBO} crystals) and half-wave plates
($\lambda/2$) in both photonic paths. In this way the initial
two-photon state is prepared in the polarization singlet state
$|\psi_s \rangle=(|HV\rangle-|VH\rangle)/\sqrt{2}$, where $H(=0)$
and $V(=1)$ are labels for horizontal and vertical polarizations
of the two photons, respectively.

In the second part of the experimental set-up (\textsf{MEMS
preparation}) the \emph{two-term} decomposition of
$\mathcal{M}_\mathcal{T}$ is physically realized by a
\emph{two-path} optical device.  A photon enters such a device
through a $50/50$ beam splitter (\textsf{BS}) and can be either
transmitted to path $1$ or reflected to path $2$. The two paths
defines two independent \emph{spatial} modes of the field.
In path $1$ a neutral-density filter ($\mathrm{\textsf{A}}_1$) is
followed by a linear polarizer (\textsf{P}) oriented horizontally
(with respect to the \textsf{BBO} crystal basis). When the photon
goes in this path, the initial singlet is reduced to $| HV
\rangle$ with probability proportional to the attenuation ratio
$a_1$ of $\mathrm{\textsf{A}}_1$ ($a
=P_\mathrm{out}/P_\mathrm{in}$). In path $2$ a second
neutral-density filter ($\mathrm{\textsf{A}}_2$) is followed by
two half wave-plates ($\lambda/{2}$) in cascade relatively
oriented at $45^{\circ}$: they work as a $90^\circ$ polarization
rotator. When the photon
 goes in path $2$, the singlet undergoes a local rotation with
probability proportional to the attenuation ratio $a_2$ of
$\mathrm{\textsf{A}}_2$.
The third and last part of the experimental set-up
(\textsf{Tomographic analysis}), consists of two tomographic
analyzers (one per photon), each  made of a quarter-wave plate
($\lambda/4$) followed by a linear polarizer (\textsf{P}). Such
analyzers permit a tomographically complete reconstruction, via a
maximum-likelihood  technique \cite{James01}, of the two-photon
state. Additionally, interference filters (\textsf{IF}) in front
of each detector ($\Delta \lambda = 5$ nm) provide for bandwidth
selection.
It should be noticed that detector
$\mathrm{\textsf{D}}_\mathrm{\textsf{A}}$ does not distinguish
which path (either $1$ or $2$) a photon comes from, thus photon
$A$ is detected in a \emph{mode-insensitive} way: This is the
simple mechanism we use to induce decoherence. In the actual
setup, a lens (not shown in Fig. 1) placed in front of detector
$\mathrm{\textsf{D}}_\mathrm{\textsf{A}}$ focusses both paths $1$
and $2$ upon the sensitive area of the detector which becomes thus
unable to distinguish between photons coming from either path $1$
or $2$ (``mode-insensitive detection'').

\begin{figure}[!hbr]
\includegraphics[angle=0,width=7truecm]{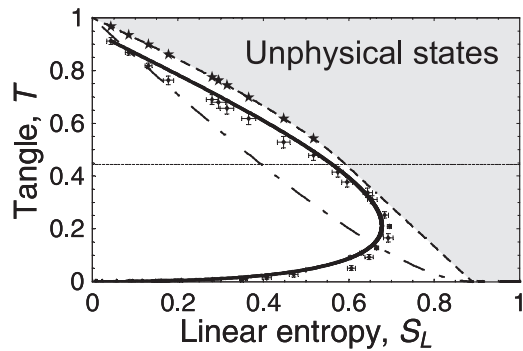}
\caption{\label{fig:B} Experimental data  and theoretical
prediction (continuous line) in the linear entropy-tangle plane.
The gray region represents unphysical states and it is bounded
from below by MEMS (dashed curve). The lower dotted-dashed curve
represents Werner states. The horizontal (dotted) line at $T =
4/9$  separates MEMS I  (above), from MEMS II (below). Stars
denote MEMS I states $\rho_\star$ that have the same linear
entropy as the measured states $\rho_\mathrm{I}^\mathrm{exp}$
(i.e., the experimental points above the line $T = 4/9$).
All measured data follow very well the theoretical curve.}
\end{figure}

Experimental results  are shown in Fig. 2 together with
theoretical predictions  in the linear entropy-tangle plane. The
agreement between theoretical predictions and measured data is
very good.
The experimentally prepared initial singlet state
$\rho_s^\mathrm{exp}$ has a fidelity \cite{Jozsa} $F(\rho_s,
\rho_s^\mathrm{exp}) = \left|\tr( \sqrt{\sqrt{\rho_s}
\rho_s^\mathrm{exp} \sqrt{\rho_s}})\right|^2 \sim 97 \% $ with the
theoretical singlet state $\rho_s$.
The continuous curve is calculated from the matrix  $\rho_c:
\rho_c = \mathcal{M}_\mathcal{T} \rho_s^\mathrm{exp}$,  and
varying $p$. It represents our theoretical prediction for the
given initially prepared state $\rho_s^\mathrm{exp}$.
If it were possible to achieve exactly $\rho_s^\mathrm{exp} =
\rho_s$, then such curve would coincide with the MEMS curve above
the horizontal (dotted) line $T = 4/9 $.
Experimental points with $T \geq 4/9$
($\rho_\mathrm{I}^\mathrm{exp}$) are obtained by varying the
neutral-density filters
$\mathrm{\textsf{A}}_1,\mathrm{\textsf{A}}_2$ in such a way that $
a_2 \geq a_1$; while
 points with $T < 4/9$ are achieved for $a_2 <a_1 $. Note that the latter points
do not represent MEMSs, but different mixed entangled states whose
density matrix is still given by $\rho_\mathrm{I}$ but with the
parameter $p$ now varying as $0 \leq p \leq 2/3$.
The average fidelity between the measured states
$\rho_\mathrm{I}^\mathrm{exp}$ an the ``target'' states
$\rho_\star$, is given by $\overline{F(\rho_\star,
\rho_\mathrm{I}^\mathrm{exp})} \sim 80 \% $.  The main reason for
its deviation from $\lesssim 100 \%$, is due to spurious,
uncontrolled birefringence in the \textsf{BS} and the prism
composing the set-up. To verify this, first we calculated the
fidelity between the  states $\rho_c(p)$ (obtained by applying the
theoretically determined map $\mathcal{T}\otimes \mathcal{I}$ to
the experimentally prepared initial singlet state
$\rho_s^\mathrm{exp}$), with the theoretical MEMS
$\rho_\mathrm{I}(p)$.  We have found $F[\rho_\mathrm{I}(p),
\rho_c(p)] \geq 97 \% $ for all $2/3 \leq p \leq 1$; thus the
value of $\overline{F} \sim 80 \% $ cannot be ascribed to the
imperfect initial singlet preparation. Second, we explicitly
measured the Mueller matrices for both the \textsf{BS} and the
prism (matrices that would be equal to the identity for ideal
non-birefringent elements) and we actually found spurious
birefringence. From such measured matrices it was possible to
determine the unwanted local unitary operation induced by these
optical elements \cite{Aiello07}. It is important to notice that
such operation does not change the position of our experimental
points in the linear entropy-tangle plane. Now, if one applies
 this unitary operation to our raw data and calculates once
again the average fidelity, the result would be $\overline{F} \sim
91 \% $. However, since this ``compensation'' of the spurious
birefringence is performed upon the measured data and not directly
on the physical setup, we felt that it was more fair to present
the uncorrected data.
\section{Discussion and conclusions}
In conclusion, we have theoretically proposed  and experimentally
tested a new, simple method to create MEMS I states of photons.
This method  can be easily generalized to generate MEMS II states,
as well. However, this task would require a slightly different
experimental setup with a \emph{three}-path linear optical device
acting only upon photon $A$  \cite{Aiello07}. In particular, we
have shown that it is possible to create a MEMS
 from a SPDC photon pair, by acting on just a \emph{single}
photon of the pair. This task could appear, at first sight,
impossible since it was recently demonstrated \cite{Ziman1} that
even the most general local operation cannot generate MEMS because
this would violate relativistic causality. However, as we
discussed in the text, our results do not contradict Ref.
\cite{Ziman1} since we obtained them via postselection operated by
coincidence measurements. The latter are possible only when causal
classical communication between detectors is permitted.
Still, the connection between relativistic causality, dichroic
(or, PDL) devices and postselection, is far from being trivial.
For example, suppose that a two-photon state is produced by an
optical setup containing \emph{local} PDL elements,  and that we
tomographically reconstruct it after coincidence measurements.
Such a reconstructed state will correctly describe the result of
any other measurement involving coincidence measurements (as,
e.g., Bell measurements), but it will \emph{fail} when describing
the result of any single-photon measurement.
We stress that this limitation is \emph{not} inherent to our
scheme, but it is shared by all optical set-ups containing PDL
elements.
\begin{acknowledgments}
We acknowledge Vladimir Bu\v{z}ek  for  useful correspondence. We
also thank Fabio Antonio Bovino for helpful suggestions. This
project is supported by FOM.
\end{acknowledgments}
%
%

%

\begin{thebibliography}{25}
\bibitem[{Not({\natexlab{b}})}]{NoteMEMS}
\bibinfo{note}{For sake of simplicity, we do not consider here type II MEMS.
  See, e.g., Ref. \cite{Peters} and references therein}.
\end{thebibliography}
%

\end{document}